\documentclass[%
reprint,
superscriptaddress,
10pt,
 amsmath,amssymb,
 aps,
]{revtex4-2}

\usepackage[utf8]{inputenc} 
\usepackage{siunitx}  
\usepackage[version=4]{mhchem}
\usepackage{chemformula}
\usepackage{graphicx}
\usepackage{braket}
\usepackage{comment}
\usepackage[rightcaption]{sidecap}
\usepackage{floatrow}
\usepackage{xcolor} % to define colors for highlighting/tracking changes
\usepackage{soul} % just for the \st command
\definecolor{darkblue}{rgb}{0, 0, 0.8}
\usepackage[colorlinks=true, breaklinks=true, linkcolor=darkblue, citecolor=darkblue, urlcolor=darkblue]{hyperref}

\newcommand*{\tx}[1]{\mathrm{#1}} % upright text in mathmode
\newcommand*{\us}{\,}             % spacing between numbers and units

\definecolor{blue2}{rgb}{0., 0.35, 1}
\definecolor{green2}{rgb}{0.1, 0.7, 0.3}

\newcommand{\ml}[1]{\textcolor{black}{#1}}

\newcommand{\env}[1]{\texttt{#1}}
\makeatletter
\newcommand*{\balancecolsandclearpage}{%
  \close@column@grid
}
\makeatother

\begin{document}
%\raggedbottom

% why plural s in "crossings"? we just have one crossing, no?
\title{Spin Reversal of a Quantum Hall Ferromagnet at a Landau Level Crossing}
% What about removing the "s"?  See e.g. https://journals.aps.org/prb/pdf/10.1103/PhysRevB.89.165313

%\author{M. Lupatini$^1$, P. Knüppel$^2$, S. Faelt$^{1,2}$, R. Winkler$^3$, M. Shayegan$^{1,4}$, A. Imamoglu$^2$, W. Wegscheider$^1$}
%\email[REVTeX Support: ]{revtex@aps.org}
%\affiliation{$^1$ Solid State Physics Laboratory, ETH Zürich, CH-8093 Zürich, Switzerland}
%\affiliation{$^2$ Institute of Quantum Electronics, ETH Zürich, CH-8093, Zürich, Switzerland}
%\affiliation{$^3$ Department of Physics, Northern Illinois University, DeKalb, Illinois 60115, USA and Materials Science Division, Argonne National Laboratory, Argonne, Illinois 60439, USA}
%\affiliation{$^4$ Department of Electrical Engineering, Princeton University, Princeton, New Jersey 08544, USA}

\author{M. Lupatini}
\affiliation{Solid State Physics Laboratory, ETH Zürich, CH-8093 Zürich, Switzerland}

\author{P. Knüppel}
\affiliation{Institute of Quantum Electronics, ETH Zürich, CH-8093, Zürich, Switzerland}

\author{S. Faelt}
\affiliation{Solid State Physics Laboratory, ETH Zürich, CH-8093 Zürich, Switzerland}
\affiliation{Institute of Quantum Electronics, ETH Zürich, CH-8093, Zürich, Switzerland}

\author{R. Winkler}
\affiliation{Department of Physics, Northern Illinois University, DeKalb, Illinois 60115, USA and Materials Science Division, Argonne National Laboratory, Argonne, Illinois 60439, USA}

\author{M. Shayegan}
\affiliation{Solid State Physics Laboratory, ETH Zürich, CH-8093 Zürich, Switzerland}
\affiliation{Department of Electrical Engineering, Princeton University, Princeton, New Jersey 08544, USA}

\author{A. Imamoglu}
\affiliation{Institute of Quantum Electronics, ETH Zürich, CH-8093, Zürich, Switzerland}

\author{W. Wegscheider}
\affiliation{Solid State Physics Laboratory, ETH Zürich, CH-8093 Zürich, Switzerland}
\date{January 2020}%

%%%%%%%%%%%%%%%%%%%%%%%%%%%%%%%%%%%%%%%%%%%%%%%%%%%%%%%%%%%%%%%%%%%%%%%%%%%%%%%%%%%%%%%%%%%%%%%%%%%%%%%%%%%%%%%%%%%
% ABSTRACT %%%%%%%%%%%%%%%%%%%%%%%%%%%%%%%%%%%%%%%%%%%%%%%%%%%%%%%%%%%%%%%%%%%%%%%%%%%%%%%%%%%%%%%%%%%%%%%%%%%%%%%%

\begin{abstract}
    % My first attempt with some alternative ideas
	When Landau levels (LLs) become degenerate near the Fermi energy in the quantum Hall regime, interaction effects can drastically modify the electronic ground state. We study the quantum Hall ferromagnet formed in a two-dimensional hole gas around the LL filling factor $\nu=1$ in the vicinity of a LL crossing in the heave-hole valence band. Cavity spectroscopy in the strong-coupling regime allows us to optically extract the two-dimensional hole gas' spin polarization. By analyzing this polarization as a function of hole density and magnetic field, we observe a spin flip of the ferromagnet. Furthermore, the depolarization away from $\nu=1$ accelerates close to the LL crossing. This is indicative of an increase in the size of Skyrmion excitations as the effective Zeeman energy vanishes at the LL crossing.
\end{abstract}

\maketitle

%%%%%%%%%%%%%%%%%%%%%%%%%%%%%%%%%%%%%%%%%%%%%%%%%%%%%%%%%%%%%%%%%%%%%%%%%%%%%%%%%%%%%%%%%%%%%%%%%%%%%%%%%%%%%%%%%%%
% INSTRUCTIONS %%%%%%%%%%%%%%%%%%%%%%%%%%%%%%%%%%%%%%%%%%%%%%%%%%%%%%%%%%%%%%%%%%%%%%%%%%%%%%%%%%%%%%%%%%%%%%%%%%%%
The quantum Hall state at the Landau level (LL) filling factor $\nu = 1$ constitutes an exciting platform for studies of many-body physics.
It has a rich phase diagram depending on the Coulomb, Zeeman and disorder potentials. 
At low disorder potential, the ground state at $\nu = 1$ is an itinerant quantum Hall ferromagnet (QHF) which can persist even in the absence of a single-particle energy gap due to strong many-body interactions \cite{Karlhede1993,Sinova2000,Girvin2000,Zhitomirsky2004,Ezawa2009}. 
Furthermore, in the absence of Zeeman coupling, a zero-temperature phase transition from a ferromagnet to a quantum Hall Skyrmion glass (QHSG) \ml{is predicted} for a large but finite Coulomb to disorder ratio, in the presence of density fluctuations \cite{Rapsch2002}. 

Previous investigations showed a high degree of spin polarization in a two-dimensional electron gas (2DEG) at $\nu = 1$ measured by different techniques such as NMR \cite{Barrett1995a,Khandelwal1998} and optical spectroscopy \cite{Aifer1996,Manfra1997}. 
In these studies, the relevance of Skyrmionic excitations in the vicinity of $\nu = 1$ was discussed to explain the fast depolarization of the system away from $\nu = 1$.
It was also pointed out that, both in GaAs 2DEGs and two-dimensional hole gases (2DHGs), Skyrmions are the lowest-lying charged excitations in the vicinity of $\nu = 1$ \cite{Karlhede1993,Barrett1995a,Schmeller1995,Aifer1996,Townsley2005,Bryja2006}.
Skyrmionic excitations rely on the interplay between Zeeman energy $E_{\text{Z}}$ and the Coulomb energy $E_{\text{C}}$. When the ratio between these two quantities $E_{\text{Z}}/E_{\text{C}}$ is small, large Skyrmions are favoured. Eventually, in an ideal system with vanishing $E_{\text{Z}}$, the Skyrmion size would become macroscopic \cite{Karlhede1993,Leadley1998a,Shukla1999}.
\ml{Vanishing $E_{\text{Z}}$ was experimentally achieved by applying hydrostatic pressure \cite{Leadley1998a} or in {Al}$_{0.13}${Ga}$_{0.87}${As}$/${GaAs} quantum wells (QWs) where the g-factor is close to zero \cite{Shukla1999}. In both cases, large Skyrmions were observed.}
Another possibility to obtain a vanishing $E_{\text{Z}}$ is to exploit LL crossings in a 2DHG, which occur at finite magnetic field due to heavy- and light-hole mixing in the valence band \cite{Schlesinger1985,Hawksworth1993,Cole1997,Winkler2003,Kubisa2003,Fischer2007a,Liu2014a,Liu2015}. In this scenario, an interesting, but so far unresolved question, is the robustness
of the QHF and the role of Skyrmions at the LL crossing.

In this Letter, we study the $\nu = 1$ QHF in a density-tunable 2DHG sample in the vicinity of a LL crossing. 
We extract the absolute value of the spin polarization using cavity spectroscopy from which we obtain a phase diagram of the $\nu = 1$ state, clearly showing a spin reversal of the ferromagnetic ground state. A narrow transition region with vanishing average polarization separates the two highly-spin-polarized phases, while the transition region broadens away from $\nu = 1$. 
By approaching the LL crossing, thereby tuning $E_{\text{Z}}$ to zero, we observe a faster depolarization of the QHF as a function of the filling factor, suggesting a diverging Skyrmion size.

\begin{comment}
\begin{figure*}[t]
	\centering
	\floatbox[{\capbeside\thisfloatsetup{capbesideposition={right,center},capbesidewidth=52mm}}]{figure}[\FBwidth]
 	{\caption{
 	(a) Sample structure. The etched region is probed on the border of the sample where the top mirror is partially removed. The cavity region is a $4\times \SI{4}{mm^2}$ square in the center of the sample. (b) White light reflection in the etched region as a function of the gate voltage. (c) Calculated electron and heavy-hole Landau level fan diagram for a sample with a 2DHG density of $4 \times 10^{10}$ $\si{cm^{-2}}$. (d) \ml{Extraction of spin polarization $P$ for a 2DHG density of $p=4.9 \times 10^{10}$ $\si{cm^{-2}}$.} The top panel shows the difference between $\sigma_-$ and $\sigma_+$ white light reflection spectra in the cavity region as a function of the filling factor. The black dots mark the center energy of the Lorentzian fits. The bottom panel shows the spin polarization extracted using Eq.~\ref{eq:spin_polarization} as a function of filling factor.}
  	\label{fig1}}
  	{\includegraphics[width=120mm]{Figure1_v2.png}}
\end{figure*}
\end{comment}
\begin{figure}[t]
	\centering
	\includegraphics[width=\columnwidth]{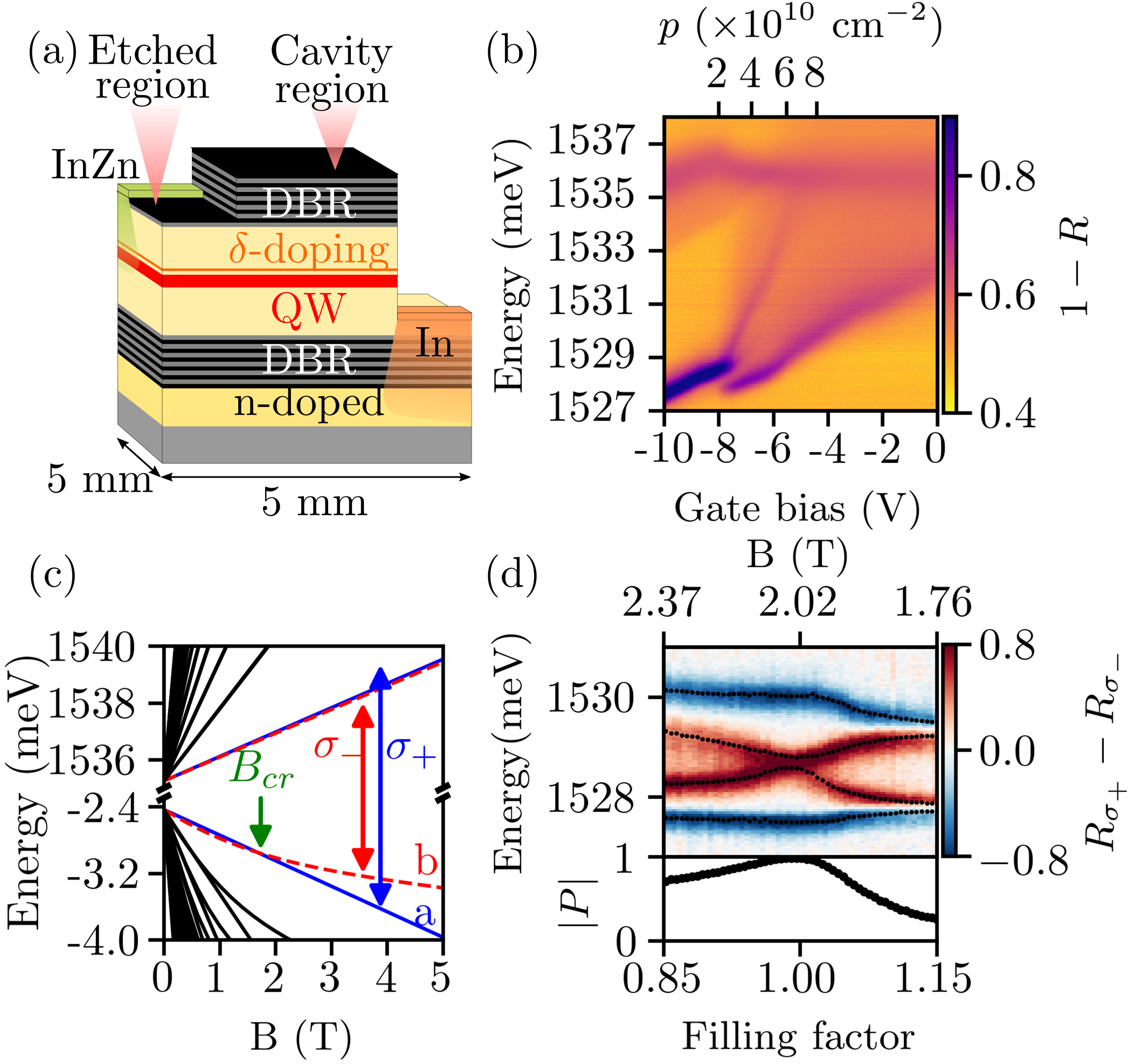}\\
	\caption{
	(a) \ml{Sample structure. The cavity region is at the center of the sample, while the etched region at the border.}
	%Sample structure. The etched region is probed on the border of the sample where the top mirror is partially removed. The cavity region is a $4\times \SI{4}{mm^2}$ square in the center of the sample. 
	(b) White light reflection in the etched region as a function of the gate voltage. (c) Calculated electron and heavy-hole Landau level fan diagram for a sample with a 2DHG density of $4 \times 10^{10}$ $\si{cm^{-2}}$. (d) \ml{Extraction of spin polarization $P$ for a 2DHG density of $p=4.9 \times 10^{10}$ $\si{cm^{-2}}$.} The top panel shows the difference between $\sigma_-$ and $\sigma_+$ white light reflection spectra in the cavity region as a function of the filling factor. The black dots mark the center energy of the Lorentzian fits. The bottom panel shows the extracted $|P|$ using Eq.~\ref{eq:spin_polarization} as a function of filling factor.}
	\label{fig1}
\end{figure} 
\ml{The sample was grown by molecular beam epitaxy on a (001) GaAs wafer, and its structure is shown in Fig.~\ref{fig1}(a).
The 2DHG is formed in a $15$-$\tx{nm}$-thick, single-side modulation-doped GaAs QW, which is embedded at the centre of a single-wavelength-long $\ce{Al_{0.21}Ga_{0.79}As}$ optical microcavity. The carbon $\delta$-doping is located $\SI{30}{nm}$ above the QW.} The top and bottom distributed Bragg reflectors (DBRs) are composed of 21 and 25 pairs of $\ce{AlAs/Al_{0.21}Ga_{0.79}As}$ layers, respectively. The microcavity has a quality factor of $Q \approx 5 \times 10^3$ which was measured by white light reflection. A silicon-doped n-layer is grown below the bottom DBR acting as a gate which allows tuning the hole density $p$ of the 2DHG.
At the border of the sample 16 DBRs pairs were etched with selective wet etching to contact the 2DHG which was achieved with $\ce{In_{0.96}Zn_{0.04}}$. A second etching step was performed to etch under the 2DHG and contact the n-type gate with In. 
The measured mobility, from a sample of the same wafer with a patterned $100$-$\tx{\mu m}$-wide Hall bar at $\SI{250}{mK}$, is about $1.14\times 10^6 \us \si{cm^2/Vs}$ at a density of $1.92 \times 10^{11} \us \si{cm^{-2}}$ obtained by Hall measurement. Applying a negative bias to the diode-like structure, the 2DHG is linearly depleted. 
The hole mobility decreases linearly with decreasing the hole density $p$, down to $0.33\times 10^6 \us \si{cm^2/Vs}$ at a density of $0.57 \times 10^{11} \us \si{cm^{-2}}$.

In a dilution refrigerator at about $\SI{70}{mK}$, the optical resonances of the 2DHG are probed by white light reflection as a function of the gate voltage in the region where the top mirror was partially removed. At large negative bias ($\leqslant \SI{-8}{V}$) the 2DHG is depleted, only the heavy-hole and light-hole exciton modes at about $\SI{1528.5}{meV}$ and $\SI{1536}{meV}$ are observed (Fig.~\ref{fig1}(b)). The shift to lower energies from $\si{-8}$ to $\SI{-10}{V}$ is attributed to the quantum-confined Stark effect (QCSE) \cite{Miller1984}.
At about $\SI{-7.9}{V}$, a low density 2DHG is formed which interacts with excitons and splits the resonance into two polaronic modes. These two modes arise from the dressing of the optically excited excitons by excitations of the heavy-hole Fermi sea, thereby forming an attractive and a repulsive exciton-polaron resonance \cite{Rapaport2000,Rapaport2001,suris_correlation_2003,BarJoseph2005,Sidler2016b,Efimkin2017,Ravets2017a}. 
The heavy-hole attractive exciton-polaron, previously referred to as the charged exciton or trion \cite{BarJoseph2005}, appears at an energy of approximately $\SI{1}{meV}$ lower than the exciton mode.  
The heavy-hole repulsive exciton-polaron is also well visible from $\SI{-5.5}{V}$ up to $\SI{-8}{V}$. Due to phase-space filling, as well as due to QCSE, the modes shift to higher energies with increasing gate voltage. We also observe the light-hole exciton dressed by the heavy-hole Fermi sea forming an attractive polaron at an energy $\sim \SI{0.5}{meV}$ lower than the corresponding (light-hole) exciton. In contrast to the heavy-hole exciton-polaron, there appears to be a continuous transition from the bare light hole exciton mode distinguishable from the polaron branch.

In the following, we will focus on the region of the sample with the microcavity where the exciton-polarons hybridize with the cavity mode to form polaron-polaritons \cite{Rapaport2000,Rapaport2001,smolka2014,Ravets2017a}. 
Given the complex behaviour of the heavy-hole LLs, we performed numerical calculations based on the $8\times 8$ Kane model \cite{Winkler2003}.  Many-body interactions were taken into account within the self-consistent Hartree approximation.  Due to mixing between the heavy- and light-hole states, the LLs are non-linearly dependent on the magnetic field $B$ (Fig.~\ref{fig1}(c)), resulting in multiple LL crossings.  In this work we study the QHF in the vicinity of the crossing between the two lowest LLs occurring at a magnetic field of approximately $B_{\text{cr}} \approx \SI{1.6}{T}$ (see green arrow in Fig.~\ref{fig1}(c)).
The LL labeled "a" is a pure heavy-hole state (spin $z$ component $S_z = +3/2$) for all values of $B$, whereas the LL labeled "b" is pure heavy-hole ($S_z = -3/2$) only in the limit $B=0$, but it acquires some light-hole character ($S_z = +1/2$) at finite $B$ (about $\SI{10}{\%}$ at $B=\SI{5}{T}$). Note that the spin $z$ component is expressed according to the "hole notation" \footnote{In the valence electron notation, the spin $z$ components of the valence band states would be the opposite of the ones specified in this Letter}.
The effective Zeeman gap $\tilde{E}_{\text{Z}}$ between the LLs a and b is very small at magnetic fields $B < B_{\text{cr}}$, \ml{ while it quickly increases for $B > B_{\text{cr}}$.}  %\pk{\st{For $B > B_{\text{cr}}$, $\tilde{E}_{\text{Z}}$ quickly increases to hundreds of $\si{\mu eV}$. }}
%\pk{\st{In Fig.~\ref{fig1}(c), the optical transitions are represented according to the selection rules for dipole transitions.}}

 To study the QHF around $\nu = 1$, we perform polarization-resolved optical spectroscopy in perpendicular magnetic fields to obtain the spin polarization of the 2DHG. The white light reflection is probed independently for $\sigma_-$ and $\sigma_+$ polarized light. In the top panel of Fig.~\ref{fig1}(d) the difference between the $\sigma_-$ (red) and $\sigma_+$ (blue) reflectivities is shown for $p = 4.9 \times 10^{10}$ $\si{cm^{-2}}$ as a function of $\nu$. 
 The polariton normal-mode splitting $\Omega$ corresponds to half the energy splitting between the upper and lower polariton at zero cavity detuning. 
 For the displayed range of filling factors, a large polariton normal-mode splitting is observed for $\sigma_+$ polarization. However, the normal-mode splitting for $\sigma_-$ vanishes at $\nu = 1$ suggesting a complete spin polarization of holes in the lowest LL spin subband. 
 Indeed, the normal-mode splitting squared $\Omega_{\sigma_{\pm}}^2$ reflects the polaron oscillator strength and therefore the available density of states of each spin subband of the first LL. This allows for a quantitative determination of the spin polarization
 \begin{equation}
 P = ( \Omega_{\sigma_+}^2 - \Omega_{\sigma_-}^2)/(\Omega_{\sigma_+}^2 + \Omega_{\sigma_-}^2)
 \label{eq:spin_polarization}
 \end{equation}
 of the 2DHG \cite{Ravets2017a,Plochocka2009}.
 
 For each filling factor, the white light reflection spectra for both circular polarizations were fitted with Lorentzian lineshapes. This allows extracting the normal-mode splitting and therefore the spin polarization according to Eq.~\ref{eq:spin_polarization}, as shown in the bottom panel of Fig.~\ref{fig1}(d). Almost complete spin polarization ($\SI{98}{\%}$) was measured at $\nu = 1$ consistent with the fact that the system is in a robust ferromagnetic state. \ml{The almost complete quench of the normal-mode splitting at $\nu = 1$ was not observed in previous cavity spectroscopy experiments on 2DEGs \cite{smolka2014,Ravets2017a}. Furthermore, the system remains almost fully polarized for $\nu \in [0.98,1.02]$ and then depolarizes on both sides of $\nu = 1$.}
 
  \begin{figure}[b]
	\centering
	\includegraphics[width=\columnwidth]{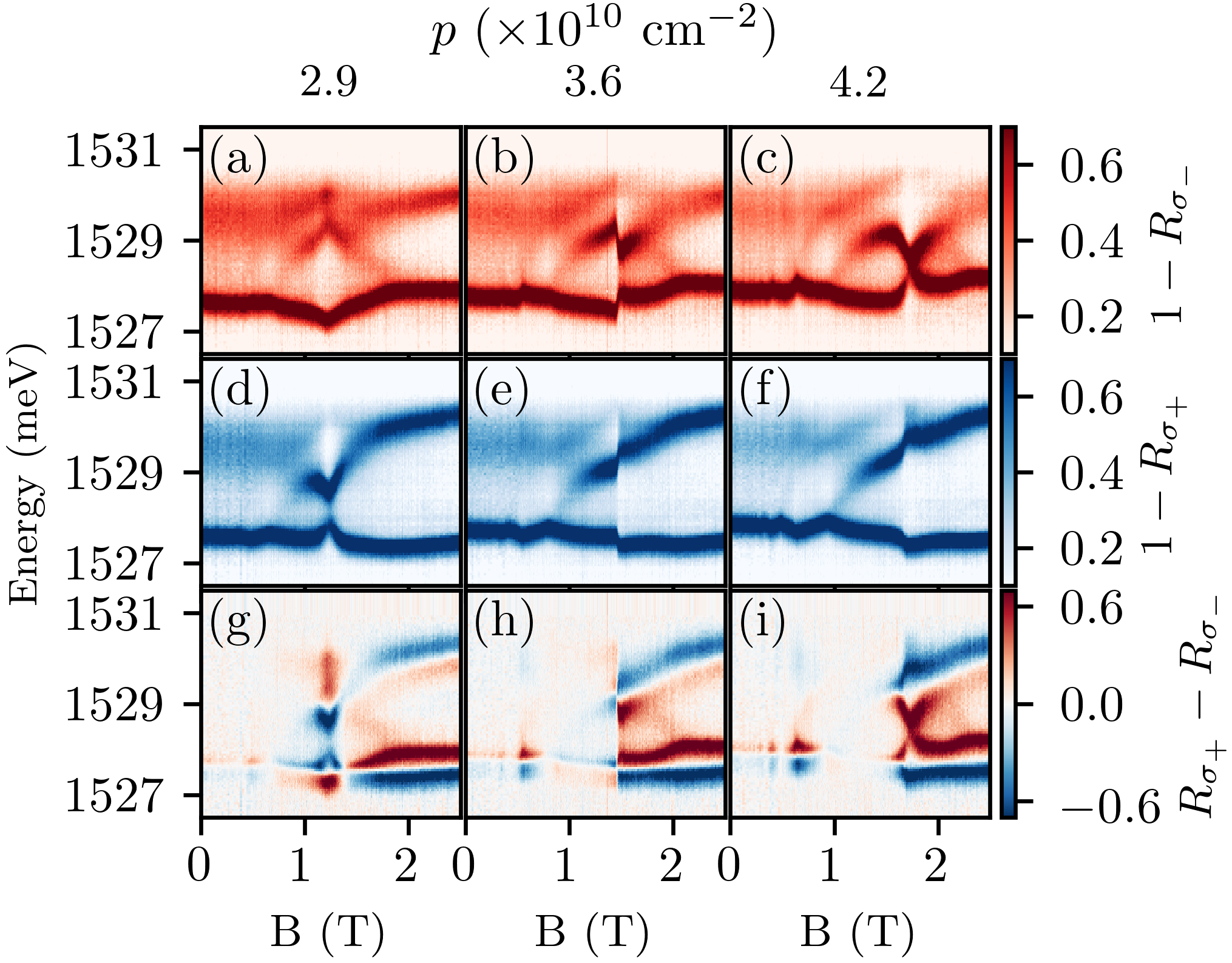}\\
	\caption{
	Circularly-polarized magnetoreflection measurements in the cavity region with $\sigma_-$ (a-c) and $\sigma_+$ (d-f) white light optical excitations. (g-i) Magnetoreflection difference between $\sigma_-$ and $\sigma_+$ optical excitations ($R_{\sigma_+} - R_{\sigma_-} = (1-R_{\sigma_-}) - ( 1-R_{\sigma_+})$). The columns correspond to three 2DHG densities as indicated on top of the figures.}
	\label{fig2}
\end{figure} 
 
 \ml{
The robustness of the QHF at $\nu = 1$ is a still source of debate. Counterposed results were shown by previous works, where in most cases only partial polarization was measured \cite{Aifer1996,Plochocka2009,smolka2014,Ravets2017a}.
\citet{Manfra1997} performed optical spectroscopy on three samples and only one of them showed full polarization. \citet{Plochocka2009} showed that the QHF is fragile and only for a temperature as low as \SI{40}{mK} a full polarization was achieved. Almost full polarization was also measured in a low-mobility 2DHG \cite{Bryja2006}.
Advancement on the understanding of the robustness of the QHF was achieved by \citet{Piot2016}. These authors showed that an optimal amount of disorder induced stabilization of the QHF in a GaAs 2DEG. Moreover, a transition from a QHF to a QHSG was proposed to occur as the ratio between the Coulomb interactions and disorder is increased leading to a lower average polarization.}

We present experimental evidence for the spin flip of the ferromagnetic ground state caused by the LL crossing in Fig.~\ref{fig2}, where white light reflection measurements as a function of the magnetic field are presented for \ml{$\sigma_-$ in Figs.~\ref{fig2}(a-c), $\sigma_+$ in Figs.~\ref{fig2}(d-f) and the difference between $\sigma_-$ and $\sigma_+$ in Figs.~\ref{fig2}(g-i)}.
At a density of $2.9 \times 10^{10}$ $\si{cm^{-2}}$ \ml{(Figs.~\ref{fig2}(a,d,g))}, the $\nu = 1$ state is at about $\SI{1.20}{T}$, where the polariton normal-mode splitting vanishes for $\sigma_+$ optical excitation \ml{(Figs.~\ref{fig2}(d))}. 
This is expected for a ferromagnetic state, where all the spins occupy the lower spin-subband of the first LL $\ket{\Uparrow}$ and therefore reduces the oscillator strength of the $\sigma_+$ exciton resonance.
In contrast, the normal-mode splitting is enhanced for $\sigma_-$ \ml{(Figs.~\ref{fig2}(a))} due to an increase of oscillator strength, due to availability of all holes for dynamic screening, and due to complete absence of phase-space filling for $\sigma_-$ exciton formation \cite{Ravets2017a}. 
At the density of $3.6 \times 10^{10}$ $\si{cm^{-2}}$ \ml{(Figs.~\ref{fig2}(b,e,h))}, a remarkably sharp transition is observed at a magnetic field of approximately $\SI{1.45}{T}$. At this density and magnetic field, the $\nu = 1$ state is exactly at the LL crossing and the polariton modes experience a sharp jump in $\Omega$ as a function of $B$. At a density of $4.2 \times 10^{10}$~$\si{cm^{-2}}$ (Figs.~\ref{fig2}(c,f,i)) $\Omega$ vanishes for $\sigma_-$ polarized light \ml{(Figs.~\ref{fig2}(c))} indicating that the ferromagnet is in the $\ket{\Downarrow}$ state. 
The magnetic field and density at which the LL crossing occurs are in good agreement with the numerical simulations (Fig.~\ref{fig1}(c)). Moreover, the numerical simulations show that for $B<B_{\text{cr}}$, the LL gap between the first and second levels is much smaller than for $B>B_{\text{cr}}$. This explains the small difference between  $\sigma_-$ and $\sigma_+$ optical excitations at $B<B_{\text{cr}}$ in Fig.~\ref{fig2}\ml{(g-i)} where a clear spin splitting is observed only at $\nu = 1$.

\begin{figure}[b]
\centering
\includegraphics[width=\columnwidth]{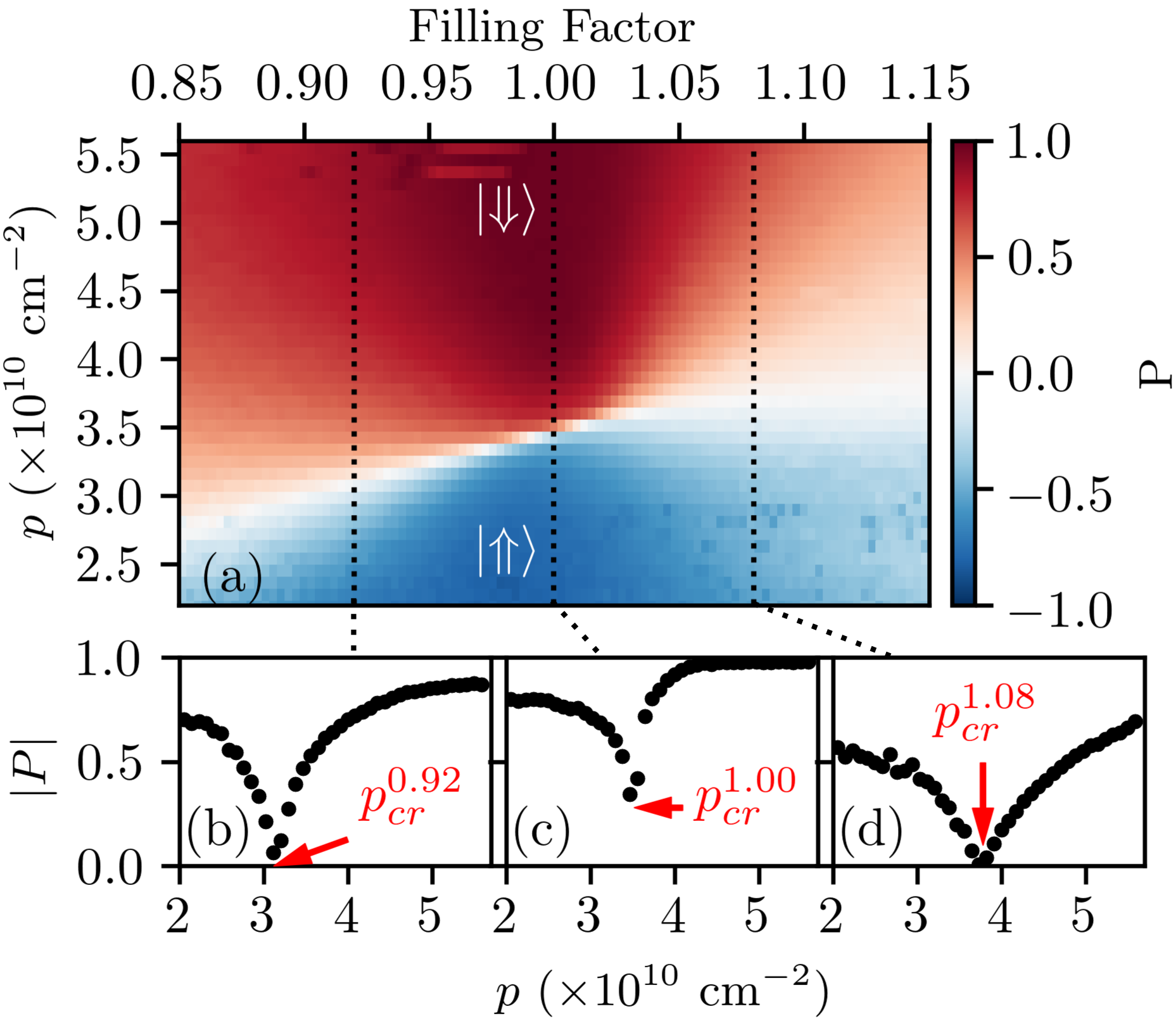}
\caption{(a) Color map of the spin polarization $\text{P}$ measured by magnetoreflection spectroscopy and extracted according to Eq. \ref{eq:spin_polarization} as a function of filling factor (x-axis) and 2DHG density (y-axis). (b, c, d) Line cuts of the absolute polarization of (a) at different filling factors: $\nu = 0.92, 1.00$ and $1.08$, respectively. 
}
\label{fig3}
\end{figure}

In Fig.~\ref{fig3}(a), additional evidence for the LL crossing is presented in a phase diagram of the 2DHG spin polarization as a function of density and filling factor. The LL crossing appears as a transition region with low polarization (shown in white) between two highly-spin-polarized regions. \ml{In the red region, holes are mainly in the $\ket{\Downarrow}$ state, while in the blue region, they are mainly in the $\ket{\Uparrow}$ state.}
We show, in Figs.~\ref{fig3}(b-d), three line cuts at constant filling factors of the absolute value $|P|$ to illustrate the evolution of the spin polarization away from $\nu = 1$.
\ml{At $\nu = 1$ and $p > p_\text{\text{cr}} \approx 3.6\times 10^{10} \us \si{cm^{-2}}$, the ground state is a robust QHF with $|P|$ reaching a maximum value of $\SI{98}{\%}$.}
After a sharp transition at the LL crossing, occurring at a density of approximately $p_{\text{cr}} \approx 3.6\times 10^{10} \us \si{cm^{-2}}$, the system repolarizes in the opposite spin state ($\ket{\Uparrow}$), up to an absolute value of spin polarization of $\SI{80}{\%}$ \ml{suggesting a weaker QHF state compared to the $p > p_\text{\text{cr}}$ regime.}

\ml{
These two distinct regimes are characterized by different $\tilde{E}_{\text{Z}}$ as inferred from the simulation in Fig.~\ref{fig1}(c). At $p < p_{cr}$, $\tilde{E}_{\text{Z}}$ is small which can lead, in the presence of disorder and density fluctuations, to charges being introduced to the system in the form of Skyrmions and Anti-Skyrmions. In this situation, a QHSG could be formed and results in a lower average polarization \cite{Rapsch2002,Piot2016}. At $p > p_{cr}$, $\tilde{E}_{\text{Z}}$ is large and the Skyrmion formation is unlikely allowing the formation of a robust QHF.
For high densities, $p \gtrsim 4.5 \times 10^{10} \si{cm^{-2}}$, the system remains polarized between $\nu \in [0.98,1.02]$. Depolarization takes place only at a sufficient deviation from $\nu=1$ where Skyrmions are numerous enough to overlap \cite{Rapsch2002,Piot2016}.
The line cuts in Figs.~\ref{fig3}(b, d) at $\nu = 0.92$ and $1.08$ show a broadening of the transition region, which indicates a lower energy gap compared to the line cut at $\nu = 1$, consistent with this interpretation. The asymmetry of the polarization at $\nu > 1$ and $\nu < 1$ is a consequence of the asymmetry of $\tilde{E}_{\text{Z}}$ around the LL crossing (see Fig.~\ref{fig1}(c)).}
\ml{Our experiments cannot rule out the possibility of nematic phases playing a role in loss of spin polarization, %\pk{due to an average effect over the different nematic phases,} 
in the vicinity of the LL crossing and contributing to the broadening of the transition region \cite{Shkolnikov2005,Padmanabhan2010,Gokmen2010,Abanin2010,Kumar2013,Feldman2016,Hossain2018,Parameswaran2019}.}
%\pk{In this case, the measured polarization would result in an average of the different nematic phases.}}
%Due to the presence of disorder, domains with different phases could lead to an averaging effect that would reduce the measured polarization.} 

In order to evaluate the importance of Skyrmions near the crossing, we studied the depolarization of the system when the filling factor is tuned away from $\nu = 1$ by applying the Skyrmion model proposed in \cite{Barrett1995a}:
\begin{equation}
P =
\begin{cases} 
P_{1}^{\mathcal{S}} \left[S \frac{2-\nu}{\nu} - (S-1) \right], & \nu > 1 \\ 
P_{1}^{\mathcal{A}} \left[\frac{1}{\nu} - (2A-1)\frac{1-\nu}{\nu} \right], & \nu < 1 
\end{cases}
\label{eq:skyrmion_model}
\end{equation}
where $\mathcal{S}$ and $\mathcal{A}$ describe the sizes of the Skyrmions and anti-Skyrmions in terms of number of holes whose spin flips upon removing or adding a flux quantum, respectively.

$P_{1}^{\mathcal{S}}$ or $P_{1}^{\mathcal{A}}$ are two fitting parameters representing the amplitude of the spin polarization in the limits $\nu \to 1^+$ and $\nu \to 1^-$, respectively. %\maybe{\st{This allows for a better fitting of the depolarization in the vicinity of $\nu = 1$ when the system is not fully polarized.}}
In a single-particle picture, $\mathcal{A}$ and $\mathcal{S}$ are equal to $1$ \cite{Barrett1995a}.
However, it has been shown in previous studies that in $\ce{GaAs}$ 2DEGs and 2DHGs, $\mathcal{S}$ and $\mathcal{A}$ are larger than $1$ near $\nu = 1$ \cite{Barrett1995a,Aifer1996,Khandelwal1998,Manfra1997,Townsley2005,Bryja2006}.
Their values depend on the ratio $E_{\text{Z}}/E_{\text{C}}$, where $E_{\text{Z}}$ promotes single-spin excitations reducing the Skyrmion size and therefore $\mathcal{S}$ and $\mathcal{A}$. In contrast, a large $E_{\text{C}}$ promotes large Skyrmions.

\begin{figure}[t]
\centering
\includegraphics[width=\columnwidth]{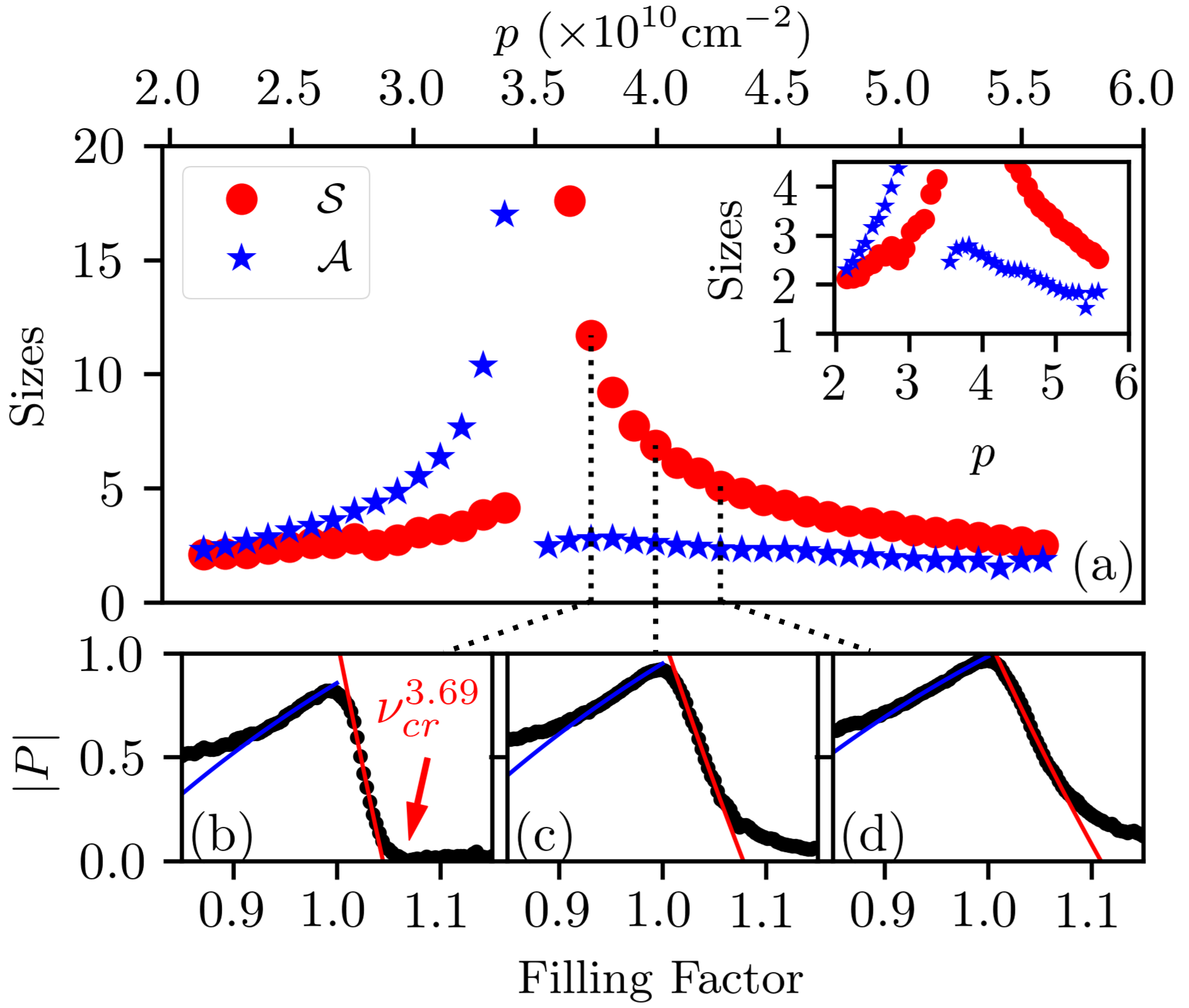}
\caption{(a) Skyrmion parameters $\mathcal{S}$ and $\mathcal{A}$ describing the depolarization around $\nu = 1$ by Eq. \ref{eq:skyrmion_model} as a function of the 2DHG density. (b, c, d) Absolute polarization as a function of the filling factor for different 2DHG densities: $p = 3.69$, $3.96$, and $4.32 \times 10^{10} \us \si{cm^{-2}}$. The red and blue lines are the fits with the Skyrmion model (Eq. \ref{eq:skyrmion_model}) to determine $\mathcal{S}$ and $\mathcal{A}$.}
\label{fig4}
\end{figure}

Figure \ref{fig4}(a) shows the magnitude of $\mathcal{A}$ and $\mathcal{S}$ obtained by fitting the data shown in Fig.~\ref{fig3}(a) using Eq. \ref{eq:skyrmion_model} as a function of $p$.
As the density \ml{approaches} $p_{\text{cr}}$, $\mathcal{A}$ and $\mathcal{S}$ diverge. 
According to particle-hole symmetry, $\mathcal{A}$ and $\mathcal{S}$ should be equal at any density \cite{Karlhede1993,Aifer1996}. However, this is not observed \ml{near} $p_{\text{cr}}$. At $p>p_{\text{cr}}$, $\mathcal{A}$ is only weakly affected by the crossing compared to $\mathcal{S}$ as can also be seen in the line cuts at $p = 3.69, 3.96$, and $4.32 \times 10^{10} \us \si{cm^{-2}}$ in Figs.~\ref{fig4}(b-d). 
For $p<p_{\text{cr}}$ the situation is the opposite, with $\mathcal{A}$ being more affected \ml{than} $\mathcal{S}$. In Fig.~\ref{fig4}(b) the crossing is clearly visible and occurs at approximately $\nu = 1.07$. The polarization drops from about $0.75$ down to $0$ in a very short range of filling factor\ml{, exhibiting a Skyrmion size $\mathcal{S} = 11.7$ while the anti-Skyrmion size $\mathcal{A} = 2.8$ remains small.}
We believe that the difference between the $\mathcal{A}$ and $\mathcal{S}$ is a direct consequence of the non-linear dependence of $\tilde{E}_{\text{Z}}$ on $B$. Indeed, a small deviation away from $B_{\text{cr}}$, especially in the case where $B > B_{\text{cr}}$, leads to a large increase of $\tilde{E}_{\text{Z}}$ \ml{(see Fig.~\ref{fig1}(c)) and a reduction of the Skyrmion and anti-Skyrmion size}. %Furthermore, this leads to a reduction of the Skyrmion and anti-Skyrmion size.

To conclude, we studied the spin polarization of the $\nu = 1$ state in the vicinity of a LL crossing occurring in 2DHGs by means of cavity spectroscopy in the strong coupling regime.
The LL crossing leads to a rich ferromagnetic phase diagram with a peculiar transition of the ground state from the $\ket{\Uparrow}$ to the $\ket{\Downarrow}$ state which is remarkably sharp at $\nu = 1$ and broadens upon tuning away from it. 
Moreover, this transition exhibits a vanishing effective Zeeman energy, which promotes large Skyrmions and therefore accelerates the depolarization of the ferromagnet upon tuning the filling factor away from $\nu = 1$.
Further studies could focus on fractional quantum Hall states in the proximity of LL crossings, enabled in 2DHGs by the richness of the valence band structure. We envision that this exciting platform will lead to new insights on quantum Hall ferromagnetism and Skyrmion physics in the limit of a vanishing Zeeman energy.

\begin{acknowledgments}
We are grateful for the support through the Swiss National Sciencce Foundation (SNF) and the NCCR QSIT (National Center of Competence in Research - Quantum Science and Technology).
M.S. acknowledges support by the DOE BES (Grant No. DE-FG02-00-ER45841), the NSF (Grant Nos. DMR 1709076 and ECCS  1906253), the Gordon and Betty Moore Foundation (Grant No. GBMF4420), and a QuantEmX travel grant from Institute for Complex Adaptive Matter and the Gordon and Betty Moore Foundation through Grant No. GBMF5305.
\end{acknowledgments}

%\newpage
\bibliography{PhD-2019_06_paper1}

% Some formatting for the supplemental
\clearpage
%\appendix % COMMAND FOR PRX APPENDIX %
%\pagebreak

\setcounter{equation}{0}
\setcounter{figure}{0}
\setcounter{table}{0}
\setcounter{page}{1}
\setcounter{footnote}{0}
\setcounter{section}{0}
\renewcommand{\theequation}{S\arabic{equation}}
\renewcommand{\thefigure}{S\arabic{figure}}

\end{document}